\documentclass[journal=naleftd,manuscript=letter,layout=twocolumn]{achemso}
\usepackage[version=3]{mhchem} 
\usepackage[T1]{fontenc}
\usepackage{multirow}
\usepackage{amsmath}
\usepackage{verbatim}

\author{M. Majka}
\affiliation[Jagiellonian University]{Marian Smoluchowski Institute of Physics, Jagiellonian University, ul. prof. Stanis\l{}awa \L{}ojasiewicza 11, 30-348 Krak\'{o}w, Poland}
\email{maciej.majka@uj.edu.pl}
\author{M. Durak-Kozica}
\author{A. Kami\'nska}
\affiliation[Jagiellonian University]{Marian Smoluchowski Institute of Physics, Jagiellonian University, ul. prof. Stanis\l{}awa \L{}ojasiewicza 11, 30-348 Krak\'{o}w, Poland}
\author{A. Opali\'nska}
\affiliation{Institute of High Pressure Physics of the Polish Science Academy, ul. Soko\l{}owska 29/37, 01-142, Warsaw, Poland}
\author{M. Szcz\k{e}ch}
\affiliation{Jerzy Haber Institute of Catalysis and Surface Chemistry Polish Academy of Sciences, ul. Niezapominajek 8, 30-239 Krak\'{o}w, Poland}
\author{E. St\k{e}pie\'n}
\affiliation[Jagiellonian University]{Marian Smoluchowski Institute of Physics, Jagiellonian University, ul. prof. Stanis\l{}awa \L{}ojasiewicza 11, 30-348 Krak\'{o}w, Poland}

\title[Subdiffusion of EVs in NTA]{The effects of subdiffusion on the NTA size measurements of extracellular vesicles in biological samples}


\begin{document}

\begin{abstract}
The interest in the extracellular vesicles (EVs) is rapidly growing as they became reliable biomarkers for many diseases. For this reason, fast and accurate techniques of EVs size characterization are the matter of utmost importance. One increasingly popular technique is the Nanoparticle Tracking Analysis (NTA), in which the diameters of EVs are calculated from their diffusion constants. The crucial assumption here is that the diffusion in NTA follows the Stokes-Einstein relation, i.e. that the Mean Square Displacement (MSD) of a particle grows linearly in time (MSD $\propto t$). However, we show that NTA violates this assumption in both artificial and biological samples, i.e. a large population of particles show a strongly sub-diffusive behaviour (MSD $\propto t^\alpha$, $0<\alpha<1$). To support this observation we present a range of experimental results for both polystyrene beads and EVs. This is also related to another problem: for the same samples there exists a huge discrepancy (by the factor of 2-4) between the sizes measured with NTA and with the direct imaging methods, such as AFM. This can be remedied by e.g. the Finite Track Length Adjustment (FTLA) method in NTA, but its applicability is limited in the biological and poly-disperse samples. On the other hand, the models of sub-diffusion rarely provide the direct relation between the size of a particle and the generalized diffusion constant. However, we solve this last problem by introducing the logarithmic model of sub-diffusion, aimed at retrieving the size data. In result, we propose a novel protocol of NTA data analysis. The accuracy of our method is on par with FTLA for small ($\simeq$200nm) particles. We apply our method to study the EVs samples and corroborate the results with AFM. Incorporating the sub-diffusive effects reduces the average measured EV diameter by $\simeq$50\% in comparison to the normal diffusion models. The remaining discrepancy between NTA and AFM can be explained with several known effects, which we also discuss in this article.
\end{abstract}

\section{Introduction}
Extracellular microvesicles (EVs) are the fragments of cell membranes generated by both prokaryotic and eukaryotic cells \cite{bib:1}. EVs diameter varies between 50 and 1000nm \cite{bib:2}. Their presence was firstly reported using hemocytometry methods in late 60's and suggested a procoagulant potential \cite{bib:3}. Neglected for the next 20 years, EVs drew back the scientists' attention in late 90s, as possible thrombotic activators \cite{bib:4}. Microscopically, they consist of cell components including a lipid bilayer, cytoplasm and ribosomal fragments \cite{bib:5}. It has been commonly accepted that EVs can be produced either by budding the cell membrane fragments (ectosomes), or by the subsequent exocytosis \cite{bib:2,bib:9,bib:10}. In blood, the main source of EVs are platelets and endothelial cells. Nevertheless, in stressing conditions (hypoxia, inflammation or hyperglycemia), also macrophages and neutrophils can release a number of EVs \cite{bib:11,bib:12}. The rapidly growing interest in EVs has a significant impact on the clinical and basic science, providing the novel potential for nanomedicine with new biomarkers, therapeutic targets and cell-to-cell communication vehicles \cite{bib:15}. The majority of clinical data, obtained from the flow cytometry studies, show the heterogeneousness of EV surface antigen profile \cite{bib:1,bib:4,bib:11,bib:12}. However, the significant progress in nanotechnology has recently allowed a systemic approach to isolation, enumeration and the characterization of EVs size, shape and molecular components \cite{bib:16, bib:FTLA2}. 

\begin{figure}
\includegraphics[width=0.98\columnwidth]{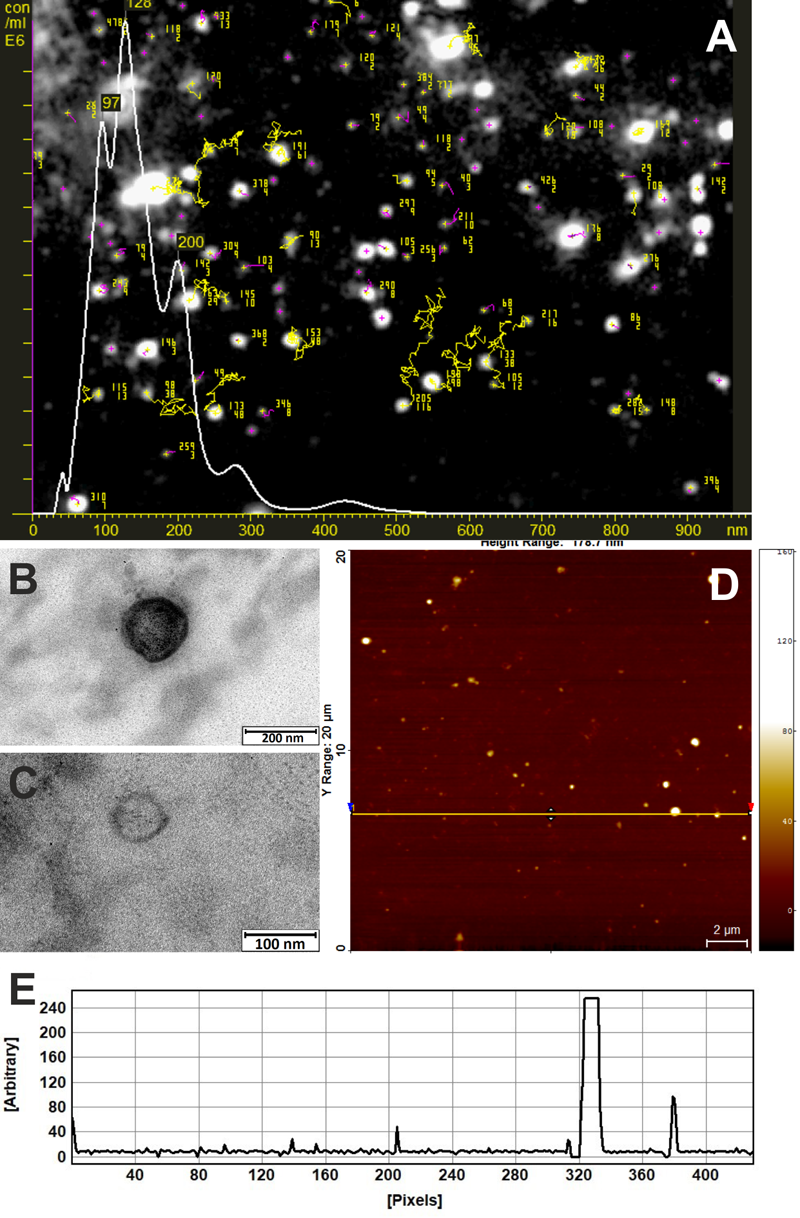}
\caption{Extracellular vesicles visualization methods. (A)  Image showing typical tracks of EVs moving under Brownian motion recorded by means of the NTA method. (B,C) Transmission Electron Microscopy (TEM) images of EVs isolated from human plasma using the ultracentifugation method. Samples were fixed with 2.5\% glutaraldehyde in 0.1M cacodylic buffer and then postfixed in 1\% osmium tetroxide solution. (D) Topography of EVs placed on a poli-L-lysine coated slide. Samples were analysed by means of Atomic Force Microscopy (AFM). The yellow line shows the cross-section, the colorimetric scale indicates the Z dimension presented as a height profile across the section line (E). \label{fig:techniques}}
\end{figure}

Present research has demonstrated that, the larger population of EVs - ectosomes - represents a rather heterogeneous population of vesicles whose size ranges from 100 to 1000 nm in diameter \cite{bib:16,bib:18}. Their size distribution can be determined with a range of methods, e.g.: scanning and transmission electron microscopy (EM) \cite{bib:5,bib:11,bib:19}, tunable resistive pulse sensing (TRPS) \cite{bib:20,bib:21,bib:22}, atomic force microscopy (AFM) \cite{bib:vanderPol1} and nanoparticle tracking analysis (NTA) \cite{bib:23,bib:vanderPol2}. Fig. \ref{fig:techniques} illustrates their application to EVs studies.

NTA is of particular interest in this article, as it is involved in 58\% of the current EVs-related research\cite{bib:58percent}. In this technique a sample of a biological fluid is observed through a microscope. A laser beam illuminates the sample from the direction perpendicular to the optical axis. The particles diffusing in the plane of the laser beam are visible thanks to the scattered light and their trajectories can be recorded (Fig. \ref{fig:techniques}A). However, the size of the particles is not concluded from the direct observation (which is constrained by the diffraction limit), but from the estimation of their diffusion constant. In the simplest approach, for each trajectory the mean square displacement (MSD) can be calculated and the Stokes-Einstein theory of diffusion predicts that MSD grows linearly in time, i.e.:
\begin{align}
\textrm{MSD}(t)=2\epsilon D_1 t && D_1=\frac{ k_BT}{3\pi\eta_0 d} \label{eq:S-E}
\end{align}
where $\epsilon$ is dimensionality ($\epsilon=2$ for NTA), $k_B$ is the Boltzmann constant, $T$ is temperature, $\eta_0$ is the viscosity of a sample and $d$ is the diameter of a particle. Equation \eqref{eq:S-E} relates the diameter $d$ of a particle to the diffusion constant $D_1$ and makes it possible to find the diameter distribution from the set of trajectories. The diffusion described by the relation \eqref{eq:S-E} is called \emph{normal}. 

NTA has entered the field of EVs research relatively recently and the evaluation of its applicability to biological samples is still in progress. Most comparative studies involving NTA focuses on its advantages over e.g. flow cytometry and dynamic light scattering (DLS)\cite{bib:vanderPol1,bib:FTLA2,bib:23}. NTA also proves to be competitive against AFM or EM, as it is cheap, fast and does not require much additional preparation of the sample that could affect the biological state of EVs\cite{bib:vanderPol1,bib:FTLA2}. Thus, it is looked at as a 'method-of-choice' in the EVs characterization with a possibly broad use in the future diagnostics. However, few studies examine whether the assumptions of NTA are satisfied in the biological context. In particular, the presence of normal diffusion is usually taken for granted. 

In this article we show that the assumption of normal diffusion is neither true in the biological samples nor, more surprisingly, for artificial mono-disperse beads. In all cases a remarkably huge population of particles has the MSD that grows sub-linearly in time. This is a strong signature of \emph{sub-diffusion}. We obtained these results consistently in several different experiments involving EVs from human plasma and artificial beads. Measurements were performed on two different NanoSight NTA analyzers (Malvern Instruments) in two separate institutes, which make us believe that the issue is inherent at least to this popular producer. Thus, the main goal of this paper is to propose the new procedure of NTA data analysis that accounts for the sub-diffusive effects.

The claim that the fundamental assumption of NTA, i.e. the relation \eqref{eq:S-E}, is not satisfied, might be surprising, as the NanoSight performs well in the industrial applications. The reason is that, by default, the NanoSight software applies the finite track length adjustment (FTLA)\cite{bib:FTLA1,bib:FTLA3} algorithm to the data. The actual implementation of FTLA in this software is not specified, but much can be inferred from the underlying study by Savney et al\cite{bib:FTLA1}. The procedure assumes that the observed distribution of sizes is the convolution of the ideal size distribution and the broadening effects from the finite length of recorded tracks. The parameters of the ideal distribution are fitted via the maximum likelihood method and this ideal distribution is provided as the final result. Two sets of assumptions are involved here. One is the form of the ideal distribution itself. For a mono-disperse sample it is expected to be a mono-peaked function. However, when the unknown number of different-sized species is involved, one cannot predict how many peaks are necessary. Indeed, FTLA performance exacerbates in the poly-disperse model systems\cite{bib:gardiner}. The other assumption is that the broadening effects follow from the normal diffusion model. As this is not satisfied, the corrections are only a crude approximation for these effects. Nevertheless, FTLA significantly improves the results for the mono-disperse model systems. However, its influence on the much more complex EVs samples is uncontrolled and, in fact, it was reported to cause artefacts\cite{bib:FTLA2}. For these reasons a more direct method is desirable.

There is also another problem related to the biological applications of NTA. When the same sample is simultaneously measured with the direct imaging methods (e.g. AFM or EM) and NTA, NTA turns out to significantly overestimate the average size. This discrepancy is observed in the study of EVs\cite{bib:NTA_Dragovic, bib:NTA_Sokolova}, but also for the coated gold nano-particles\cite{bib:nanoparticles} and, to a minor extend, model polystyrene beads\cite{bib:vanderPol2}. Our own measurements of EVs from human plasma, which we discuss further in this article, indicate that the difference between AFM and NTA can be of a factor 3-4 for normal diffusion models and about 2 for FTLA.

A few effects contributing to this difference are already recognized. One important reason is that  NTA has low sensitivity to the small EVs ($d<50$nm), which cuts out the low-end of the size distribution\cite{bib:vanderPol1,bib:NTA_Dragovic}. In our AFM study, excluding all the particles smaller than 50nm increases the mean EV diameter by roughly 34\%-40\% (see Table \ref{tab:d}). Another contribution comes from the fact that NTA measures the hydrodynamic radius, which might appear larger than the inelastic core measured in AFM and EM. This discrepancy is well-documented for the latex and polystyrene beads in monodisperse solutions, studied by both NTA and DLS\cite{bib:filipe,bib:boyd,bib:hoo}. However, to see this effect in AFM one must usually reduce the widening due to the tip convolution effect (algorithmically or by using a properly sharp tip)\cite{bib:convolution}, as it might easily dominate the results. In general, the hydrodynamic widening overestimates the size by 50\% for very small particles ($d\simeq 16$nm)\cite{bib:filipe,bib:hoo}, by 30\% for $d\simeq 50$nm, by 5-20\% for 100nm particles \cite{bib:boyd,bib:filipe,bib:hoo} and for $d>400$nm the effect is negligible\cite{bib:filipe}. There are few systematic studies on this effect in the polydisperse systems, but from the work of Filipe et al. one can conclude that the difference does not exceed 30\%\cite{bib:filipe}. Finally, NTA is influenced by the lengths of recorder trajectories, e.g. the 100nm particles can be measured as 15-25\% larger when the extremely short (less than 5 points) trajectories are used\cite{bib:gardiner}. On the other hand, the binding to a surface and tip-induced deformation might increase the apparent size of EVs in AFM. For a sphere with radius $r_s$ which is flatten into the disc with radius $r_d$, but conserves its area (so $4\pi r_s^2=2\pi r_d^2$), $r_s$ is at worst 30\% smaller than $r_d$. The deformations can be minimized by applying the 'tapping' or 'non-contact' mode in AFM. Summarizing all of these effects, the size of particles measured in NTA should be roughly 1.5-2.5 times larger than the one resulting from AFM. Therefore, the sub-diffusive model for NTA should predict the EV size that fits into this range.
  
\begin{figure}
 \includegraphics[width=0.98\columnwidth]{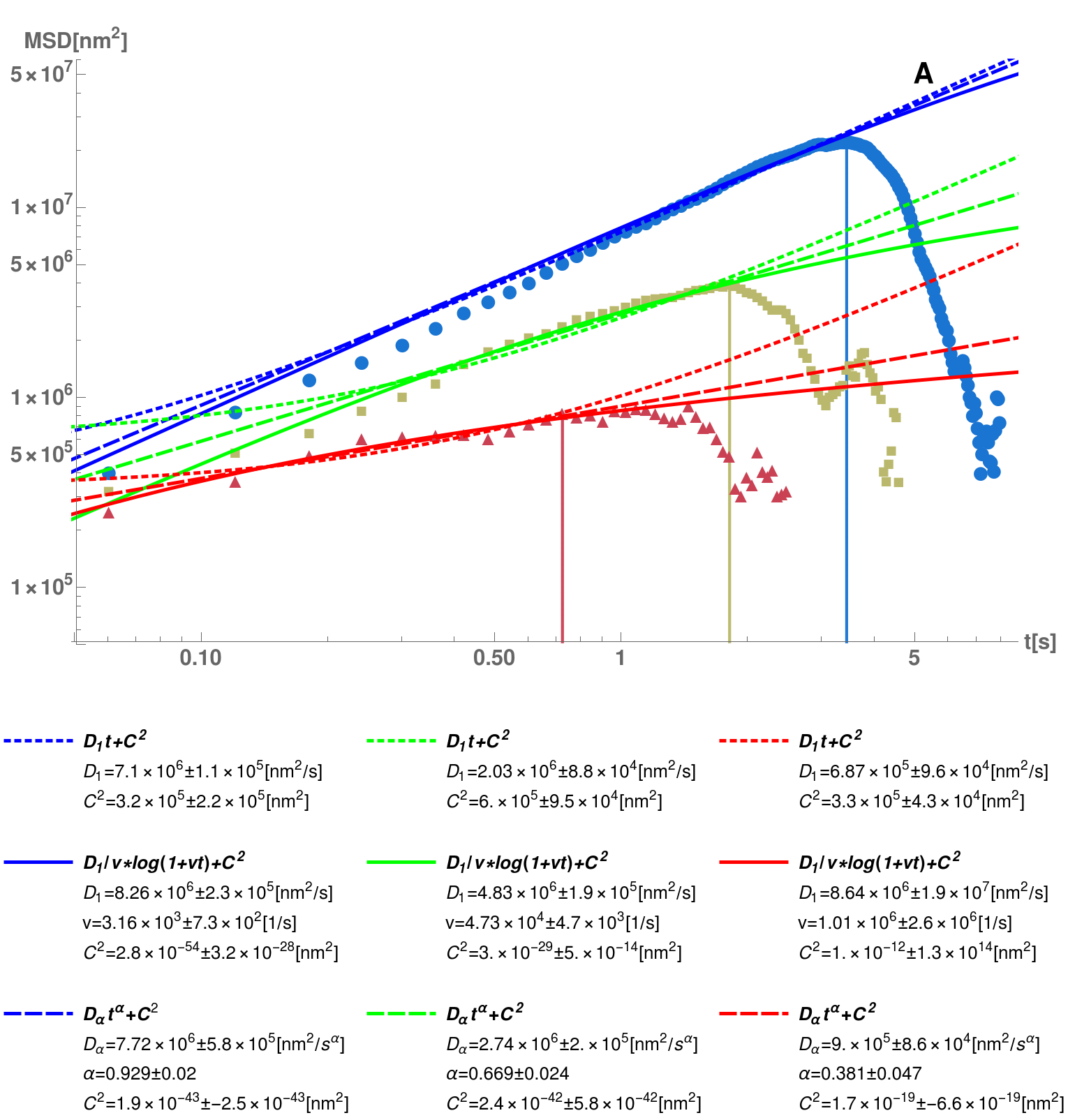}
 \includegraphics[width=0.98\columnwidth]{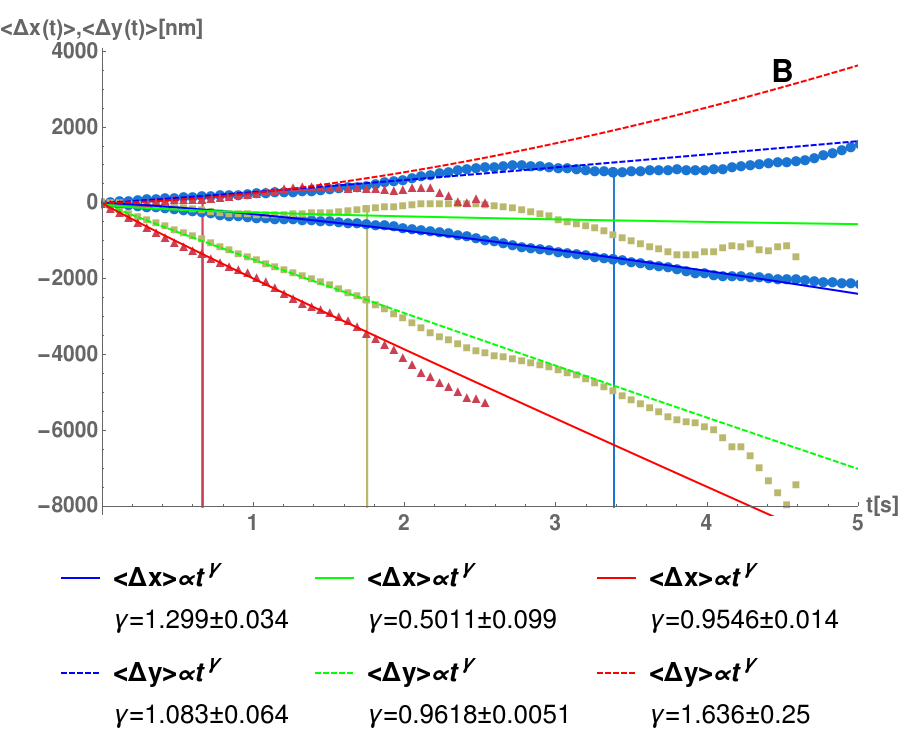}
 \caption{\textbf{A:} $\textrm{MSD}(t)$ plots for three exemplary EVs trajectories, representing normal diffusion (blue circles), intermediate sub-diffusion (green squares) and strong sub-diffusion (red triangles). The vertical lines show the cut-off time taken into analysis. The data are fitted with the linear model \eqref{eq:mod1} (short-dashed lines), power-law model \eqref{eq:mod2} (dashed lines) and logarithmic model \eqref{eq:mod3} (solid lines). For a well pronounced sub-diffusion the linear model clearly deviates from the data (red and green plots), leading to the underestimation of diffusion constant. \textbf{B:} The mean position increment $<\Delta x(t)>$ and $<\Delta y(t)>$ for each trajectory from panel B, showing the influence of drift. The increments are fitted with the power-law model $\propto t^\gamma$ to test for linearity.\label{fig:examples}}
 \end{figure}

The NTA measurements are based on the observation of the diffusive motion, which might be affected by a plethora of effects. For both mono- and poly-disperse samples these factors are e.g. confinement, molecular crowding, inhomogeneous distribution of particles and interactions between particles (electrostatic, hydrodynamic etc.). Poly-dispersity additionally introduces the unspecific excluded volume interactions\cite{bib:marenduzzo, bib:majka} and EVs might also bind biochemically with other EVs and proteins \cite{bib:12, bib:proteins}. All of these aspects violate the assumptions of the Stokes-Einstein theory\cite{bib:dybiec1,bib:dybiec2,bib:sokolov} \eqref{eq:S-E}, which describes the freely diffusing particle in an uncorrelated, molecularly homogeneous environment. Another issue might be the design of the NTA instrument itself, e.g. it might induce temperature gradients, counter-sedimentation flows and the field of view might be distorted. In what follows, the experimentally measured $\textrm{MSD}$s can deviate significantly from the linear dependence $\textrm{MSD}\propto t$. This can be seen in  Fig. \ref{fig:examples}A, where we show $\textrm{MSD}(t)$ for three exemplary trajectories from our measurements. For the short-to-intermediate time $\textrm{MSD}(t)$ resembles the free diffusion behavior, but the slope of the data on the log-log plot is clearly sub-linear, i.e. $\textrm{MSD}(t)\propto t^\alpha$, where $0<\alpha<1$. This is characteristic for the sub-diffusion. From Fig. \ref{fig:examples}A one can see that fitting the normal diffusion model, i.e. $\textrm{MSD}(t)=2\epsilon D_1t+C^2$ (where $C^2$ accounts for the localization noise, see Berglund\cite{bib:berglund}), to the sub-diffusive data leads to an underestimated $D_1$ and, in result, an overestimated $d$. The free-diffusion regime is usually culminated with a local maximum, after which $\textrm{MSD}(t)$ decreases and fluctuates strongly. This behaviour usually indicates that the boundary effects become dominant.
 
 The sub-diffusion (or, more generally, \emph{anomalous} diffusion) is a phenomenon ubiquitously observed in the biological systems\cite{bib:review,bib:weiss,bib:yeast,bib:membrane,bib:chromatin}, which gains more and more recognition in the bio- and life- sciences\cite{bib:review2}. This is particularly important since the sub-diffusion affects the efficiency of transport in the molecularly crowded environment, especially cytoplasm\cite{bib:review,bib:weiss,bib:yeast}.  Microscopically, sub-diffusion is usually caused by the geometrical confinement, trapping events that intervene with the normal diffusion or by the highly inhomogeneous viscosity of a system\cite{bib:review}. The mathematical models of sub-diffusion are numerous \cite{bib:dybiec1,bib:dybiec2,bib:sokolov,bib:sub_diff_rev,bib:metzler,bib:kou,bib:goychuk}. In particular, three classes stand out, i.e. Fractional Brownian Motion\cite{bib:kou}, Continuous Time Random Walks\cite{bib:metzler} and Scaled Brownian Motion\cite{bib:scaled1,bib:scaled2,bib:scaled3}. The former two are capable of ergodicity breaking\cite{bib:scaled3,bib:ergodic1,bib:ergodic2,bib:ergodic3}, which can be examined in the single particle tracking experiments. However, most of these models are preoccupied with obtaining the anomalous exponent $\alpha$, i.e. reproducing the asymptotic dependence:
 \begin{equation}
 \textrm{MSD}(t)=2\epsilon D_\alpha t^{\alpha} \label{eq:anomalous}
 \end{equation}
where the generalized diffusion constant $D_\alpha$ is usually assumed as a fitable parameter. Although \eqref{eq:anomalous} and \eqref{eq:S-E} seem deceptively similar, they cannot be applied in the same manner. While $D_1$ is straightforwardly related to $\eta_0$ and $d$, there is no such simple dependence for $D_{\alpha}$. This can be seen from the dimensional analysis, i.e. the units of $D_1$ and $D_{\alpha}$ are $m^2/s$ and $m^2/s^{\alpha}$, respectively. Thus, one might interpret that there is an additional time scale $\tau$ involved in the sub-diffusion, such that $D_\alpha=D_1/\tau^{\alpha-1}$. However, $\tau$ and $D_1$ cannot be simultaneously determined from the experimental data via simple fitting, i.e. if one fits MSD with $D_1 t^{\alpha}/\tau^{\alpha-1}$ model, it results in any pair of $D_1$ and $\tau$ that can be recombined into $D_{\alpha}$ corresponding to the direct fitting of model \eqref{eq:anomalous}. In what follows, it is also not possible to measure the distribution of diameters $d$ with the aid of the model \eqref{eq:anomalous}. Nevertheless, since $\alpha$ is a very convenient measure of anomalousness, we will frequently resort to it. However, for the purpose of application in NTA we must consider an alternative approach. 

It should be emphasized that while the power-law model \eqref{eq:anomalous} is a popular description of sub-diffusion, it is mainly phenomenological. In fact, sub-diffusion is always a composition of multiple microscopic processes, as indicated by the Mori-Zwanzig type models\cite{bib:kou} or by the ingenious visco-elastic approach, proposed by Goychuk\cite{bib:goychuk}. Thus, different models of similar behavior can be justified on this basis. To remedy the problem of physical interpretation for $D_{\alpha}$, we first attempted to use the modified power-law model: $\textrm{MSD}(t)=2\epsilon D_1t/(1+\frac{t}{\tau})^{1-\alpha}$, which asymptotically ($t\gg 0$) behaves like \eqref{eq:anomalous} and separates $D_1$ and $\tau$. However, fitting this model to MSD data did not result in the stable values of $D_1$ and $\tau$, similarly as in the case of \eqref{eq:anomalous}. Most probably this is because NTA do not provide enough information on the $t\to0$ behavior, for which these two models mainly differ. 

Our solution is provided by the logarithmic model, based on the following, qualitative reasoning. In the dispersed colloidal solution there is a certain average distance between particles. At the short time scale, a particle diffuses freely, but over a longer period, as $\sqrt{\textrm{MSD}(t)}$ becomes comparable to this average length-scale, a particle is more likely to encounter obstacles (i.e. other particles). Thus, the viscosity of a system 'perceived' by this particle effectively grows in time. This can also happen if the particles are systematically carried by a drift into the denser region (i.e. near a barrier). In order to account for these effects, we introduce the time dependent viscosity $\eta(t)$. In the lowest order, it can be approximated as $\eta(t)\simeq(1+vt)\eta_0$, where $v$ is some effective speed of viscosity change. Assuming that the short-time diffusion is still normal, the increment of MSD over some $dt$ reads:
\begin{equation}
d\textrm{MSD}(t)=\frac{2 \epsilon kT}{3\pi \eta(t)d}dt\simeq\frac{2\epsilon D_1}{1+vt}dt
\end{equation}
One can integrate this expression over time, to obtain:
\begin{equation}
\textrm{MSD}(t)=\int_0^tds \frac{2\epsilon D_1}{1+vs}=\frac{2\epsilon D_1}{v}\ln (1+vt) \label{eq:MSD_log}
\end{equation}
The logarithmic model can mimic the power-law behavior over several orders of magnitude, when tuned properly. In fact, it can be seen as an example of the ultra-slow scaled Brownian motion\cite{bib:scaled1,bib:scaled2} and a similar result is encountered in e.g. the granular gas\cite{bib:scaled3}. This model preserves the standard interpretation of $D_1$ and it keeps $D_1$ and $v$ separated enough to allow reliable and repetitive MSD fits. One might also check that in this model $\lim_{t\to0} \textrm{MSD}(t)/t=2\epsilon D_1$, so it predicts the non-zero diffusivity for every $t$, as desired. It also leads to the normal diffusion for $v=0$, i.e. $\lim_{v\to 0}\textrm{MSD}(t)=2\epsilon D_1 t$. The quality of fits provided by this model is comparable to the power-law model.
 
 
 \section{Samples and measurements}
For the purpose of this work we gathered the MSD data from NTA for the polystyrene (PS) beads with diameters 203 and 453 nm (Thermo Scientific) and EVs extracted from blood samples. Two series of experiments were carried out: one, involving EVs and PS beads, at the Institute of High Pressure Physics (Polish Academy of Science, Warsaw) and the other, involving only PS beads, but at several concentrations, at the Jerzy Haber Institute of Catalysis and Surface Chemistry (Polish Academy of Science, Krakow). The measurements were performed by two different teams and involved two different NTA instruments.
 
 \textbf{PS$_K$: NTA measurements of PS beads in Krakow}. The 1\% aqua solution of the 203 nm PS beads (Thermo Scientific) was diluted $10^3$ and $10^4$ times. The 1\%  solution of 453 nm PS beads was diluted $10^3$, $2\times10^3$ and $5\times10^3$ times. For each concentration and size the NTA measurements were repeated three times. For $10^3$ dilution runs lasted 25s and 60s for the others. The experiments were performed with NanoSight NS500 instrument (Malvern Instruments Ltd. United Kingdom) equipped with 405 nm laser, at a room temperature of $20-24^\circ$C (varying between measurements). The shutter and gain were adjusted manually. The NTA 3.2 (dev build 3.2.16) software was used for capturing the data as well as the preliminary and FTLA analysis.
 
 \textbf{PS$_W$: NTA measurements of PS beads in Warsaw}. The 1\% solutions of 203 nm and 453 nm polystyrene beads (Thermo Scientific) were diluted $7\times10^3$ and $5\times10^3$ times, respectively. The NTA measurement lasted 90 s for bigger particles and 60 s for the smaller, each run was repeated three times. The temperature read $29^\circ$C. The NanoSight NS500 instrument (Malvern Instruments Ltd. United Kingdom) equipped with 405 nm laser was used. The shutter and gain were adjusted manually and the NTA 2.3 (build 0025) software was applied to capture the data as well as for the initial and FTLA analysis.
 
\textbf{EVs: blood donors}. Samples from 3 healthy donors (coded further as P1, P2 and P3) were used for this study. The samples has been collected in the conjunction with our previous study on diabetics\cite{bib:25}.

\textbf{Ethics}. Bioethical Committee at Jagiellonian University Medical College (JUMC) accepted all project's protocols and forms, including an information for patients form and a consent form for participation in a research study. The permission No. KBET/206/B/2013 is valid until 31st of December 2017.

\textbf{Blood collection and platelet poor plasma (PPP) preparation}. All blood samples were drawn at the same time of the day (between 08:00 and 10:00 am) with venipuncture with > 21-gauge needle in the antecubital vein following the application of a light tourniquet. Citrate blood was centrifuged twice at 2500 g for 15 min to obtain PPP. The plasma samples were aliquoted and frozen at $-80^\circ$C until further analysis. Before measurements samples were thawed in a $37^\circ$C water bath and vortexed for 30 s.

\textbf{Nanoparticle Tracking Analysis (NTA)  of plasma EVs}. To avoid EVs aggregation, all samples were diluted 100x in HEPES buffer (10 mM Hepes/NaOH, 140 mM NaCl, 2.5 mM CaCl$_2$, pH 7.4). The NTA measurement was performed with a NanoSight NS500 instrument (Malvern Instruments Ltd, United Kingdom), equipped with a sample chamber with a 405-nm laser. The assay was performed at room temperature $23.3\pm0.1^\circ$C. For each donor (P1-P3) the measurements has been carried out 3 times. The samples were measured for 30s with the manual shutter and gain adjustments in advanced settings. The NTA 2.3 Build 0025 software was used for capturing of the data as well as their preliminary and FTLA analysis. 

\textbf{Atomic Force Microscopy (AFM) analysis of plasma EVs}. Poly-L-Lysine Slides from Thermo Scientific (cat no J2800AMNZ) were previously cleaved into $1\times1$cm plates, rinsed both sides with distilled water and cleansed by compressed air. Samples with EVs were spread on glass slide and incubated for 1 hour in a humid chamber at room temperature. Then slides were rinsed three times by gently dipping in PBS (pH=7.4) and fixed in 2.5\% glutaraldehyde in PBS for 30 minutes at room temperature. After that, slides were rinsed and analyzed by means of the AFM technique. To determine the size distribution of EVs, $20\times20\mu$m topographical images of the samples were performed by Atomic Force Microscopy - Nanoscope IIIa Multimode-SPM (Veeco Instruments, Santa Barbara, CA, USA) in contact mode. AFM images were recorded in liquid (PBS) using fluid chamber and imaging conditions included a 0.4 Hz scan rate, 256 points collected per line (pixel resolution). As a probe, non-conductive pyramidal silicon nitride tip (MLCT, Bruker) with a resonance frequency 10-20 kHz, nominal spring constant 0.01 N/m and with a radius 100 nm was used. The topographical analysis of AFM images was carried out using SPIP software version 6.5.2 (Image Metrology A/S, Horsholm, Denmark). In this software the \emph{Particle and Pore Analysis} mode was used to asses the EVs diameters. An exemplary image is shown in Fig. \ref{fig:techniques}D and E. Image features that were brighter (higher) than the background were recognized as particles, while the darker (lower) were identified as pores. The detection threshold was set to 0.2 nm. The algorithm identifies the area occupied by a particle and calculates the diameter of a disc with the same area. This results is provided as an EV's diameter. Finally, the "Post processing" was applied to classify EVs and the size distribution histograms were prepared. Because the resolution of AFM images performed in contact mode depends on the radius and shape of the tip (the tip convolution effect) and the tip used in this study was relatively wide, we also decided to recalculate the size of observed particles basing on correction method from Engel et al.\cite{bib:26}.

\section{NTA data analysis}\label{sec:analysis}
In the preliminary stage of analysis the NTA software identifies the trajectories of observed particles and this raw data are available to the user (\emph{alltracks} files).  Each trajectory is given as a sequence of positions $\vec{r}_i=(x_i,y_i)$ (in pixels) on the consecutive frames indexed by $i$, captured by the camera. Thanks to the \emph{frame-rate} and \emph{calibration parameter} provided by the software it is possible to recalculate these data into the actual physical units, i.e. nanometers and seconds. NTA software is also able to distinguish the valid trajectories from the artefacts on the basis of combined length and scattered light intensity criteria. These trajectories are labeled as \emph{included in the distribution} in the output \emph{.csv} files. We restrict our analysis solely to these trajectories and obtain the size distributions with FTLA algorithm for further comparison. However, the NanoSight software gives us no control over the truncation of the trajectories and drift treatment. Since our goal is to explicate the difference between the Stokes-Einstein and sub-diffusive models, we decided to perform the MSD analysis on our own, maintaining the full control over the data processing. 

Our first goal is to determine the MSD function individually, for each trajectory. In the first step of our analysis we introduce the increments :
\begin{equation}
\Delta \vec{r}_i(n)=\vec{r}_{i+n}-\vec{r}_i
\end{equation}
where $\Delta\vec{r_i}(n)=(\Delta x_i(n),\Delta y_i (n))$ and calculate their average value:
\begin{equation}
<\Delta \vec{r}_n>=\frac{1}{N-n}\sum_{i=1}^{N-n}(\vec{r}_{i+n}-\vec{r}_i)
\end{equation} 
$N$ is the maximal number of frames on which a particular molecule is recognized. Having found the increments, we can calculate the $\textrm{MSD}(n)$, which reads:
\begin{equation}
\textrm{MSD}(n)=\frac{1}{N-n}\sum_{i=1}^{N-n}(\Delta\vec{r}_{i}(n)-<\Delta \vec{r}_n>)^2
\end{equation}
One should note that ${<\Delta \vec{r}_n>}$  is also the measure of the average drift, so our definition removes the influence of a flow along a trajectory. In the Fig. \ref{fig:examples}A, we present the log-log plots of MSD for three exemplary trajectories rescaled to the actual physical units, i.e. nanometers and seconds. These plots show several features which are representative for our data. In general, MSD consists of an initial period of the linear growth on the log-log plot up to the local maximum, followed by the interval of decrease and strong fluctuations. Most of the free diffusion models, normal or sub-diffusion, cannot describe the decrease in MSD, which usually indicates the presence of boundaries in the system\cite{bib:sub_diff_rev}. One possible exception is the scaled Brownian motion \cite{bib:scaled1,bib:scaled2}, which has a tendency to form a local maximum in MSD, but even in this case the effect is most pronounced in the confinement. Thus, we conclude that the post-maximum behavior is induced by the boundaries. For this reason, we must propose the criteria to exclude it from the further analysis based on the free diffusion models.

The features we just described are most representative for the long ($N>9$) trajectories. Unfortunately, it is well known that while NTA measures a massive number of trajectories, there is no control over their length and, usually, only a mere $10-15\%$ of them has the length of $N\ge5$. This is particularly sever in the EVs samples, in which the number of trajectories is generally low. Due to the poor statistics, the shortest trajectories have very noisy MSD plots\cite{bib:saxton}, which might significantly deviate from what we described in the previous section. Additionally, the minimal trajectory length can also significantly broaden the diameter distribution \cite{bib:gardiner,bib:saxton}. In our case we face the same problem, but, in addition, we want to analyze only those trajectories for which the initial interval of MSD at least vaguely resemble the free-diffusion behavior. This decreases the number of trajectories even further. Thus, for the purpose of our study, a trajectory is accepted if its first three MSD points satisfy $\textrm{MSD}(1)<\textrm{MSD}(2)<\textrm{MSD}(3)$. This ensures the existence of a minimal, strictly growing sequence. This also means that we use trajectories of $N=4$ at least, which is still close to the $N=5$, recommended by Gardiner. For donors P1, P2 and P3 the NanoSight software has admitted respectively 919, 679 and 366 trajectories of which 73\%, 62\% and 58\% has been accepted according to our criterion. The aforementioned problems are incomparably less severe for the PS beads, for which NTA provided usually between 1500 and 6000 relevant trajectories of which 60-80\% passed our criteria.

We also had to decide on how many points of an individual MSD sequence are accounted for the free-diffusion behavior, i.e. we had to choose the truncation time $T$. Since we expect to encounter a local maximum in MSD, we propose that $T$ is equal to the first $n$ violating $\textrm{MSD}(n+2)>\textrm{MSD}(n)$. This condition can detect the systematic decrease in MSD, which usually occurs after the local maximum. However, it also allows some accidental detours from the strictly growing character of MSD (e.g. $\textrm{MSD}(n+1)<\textrm{MSD}(n)$ while $\textrm{MSD}(n+2)>\textrm{MSD}(n)$). The random variability of MSD must be expected, especially for short trajectories, since the sample size from which an individual value of $\textrm{MSD}(n)$ is calculated reads $N-n$\cite{bib:saxton}. This number is usually low (especially for EVs) and decreases with subsequent $n$.

\begin{figure}
\includegraphics[width=0.98\columnwidth]{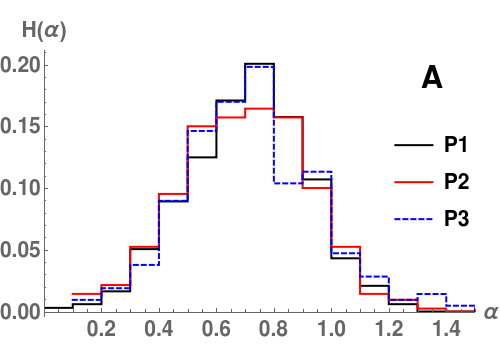}
\includegraphics[width=0.98\columnwidth]{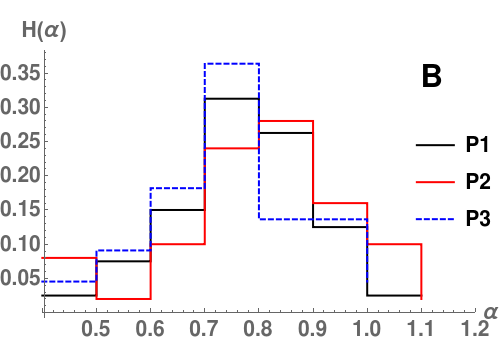}
\includegraphics[width=0.98\columnwidth]{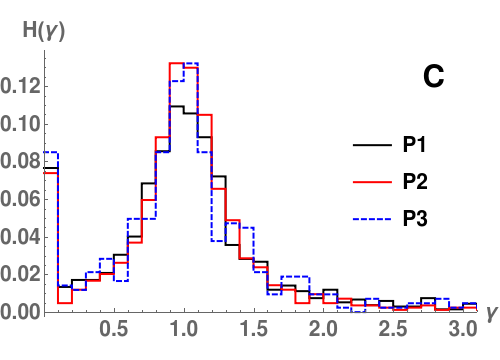}
\caption{Anomalous exponent $\alpha$ in EVs samples: \textbf{A}: The histogram of $\alpha$ resulting from the fitting of all accepted MSD data (see text) with the power-law model. For each donor the histogram has a maximum near $\alpha\simeq0.7$, indicating the sub-diffusive behaviour. \textbf{B}: The histogram of $\alpha$ obtained only for long ($N>50$) trajectories. The maximum shifts towards $\alpha\simeq 0.8$.  \textbf{C}: The joint histogram of exponent $\gamma$ resulting from fitting the average increments $<\Delta x_n>$ and $<\Delta y_n>$ with the power law model $\propto t^\gamma$. For all donors the distributions have distinct peaks at $\gamma\simeq 1$ and $\gamma\simeq 0$. \label{fig:alphas}}
\end{figure}

\begin{figure}
\includegraphics[width=0.98\columnwidth]{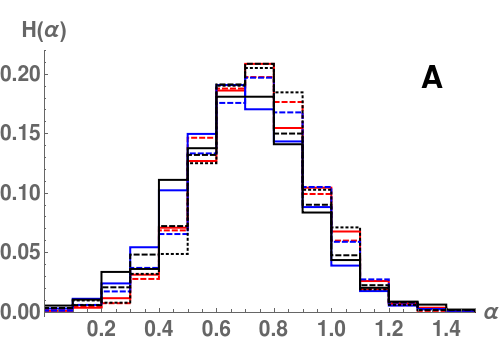}
\includegraphics[width=0.98\columnwidth]{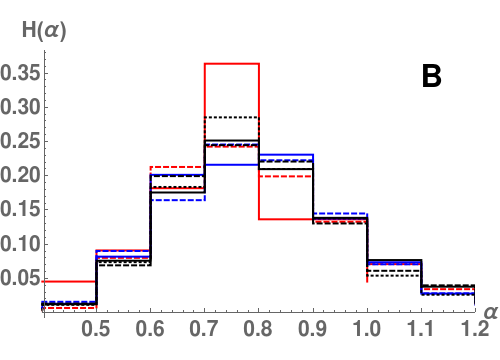}
\includegraphics[width=0.98\columnwidth]{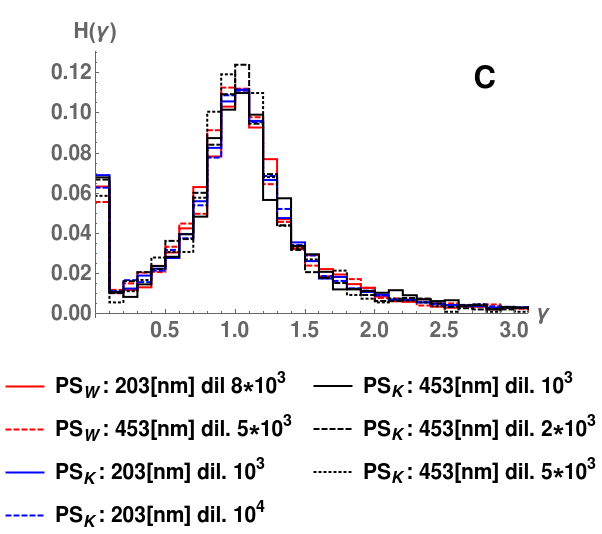}
\caption{Anomalous exponent $\alpha$ in the samples of polystyrene beads. PS$_W$ - beads measured in Warsaw, PS$_K$ - beads measured in Krakow, dil. - sample dilution. \textbf{A}: The histogram of $\alpha$ calculated for all accepted trajectories \textbf{B}: The histogram of $\alpha$ for long ($N>50$) trajectories. \textbf{C}: See caption of Fig. \ref{fig:alphas}C. \label{fig:b_alphas}}
\end{figure}

\begin{table*}
\begin{tabular}{c|c|c|c|c|c|c|c|c}
\hline
Id.  & $C_0\times 10^3$ & $d$[nm]  & $\mathcal{N}_{all}$ & $<\alpha>$ & SD & $\mathcal{N}_{N>50}$ & $<\alpha>_{N>50}$ & SD$_{N>50}$  \\
\hline
 EV & -  & - & 673 & 0.71 & 0.22 & 80 & 0.77 & 0.13  \\
EV  & - & - & 420 & 0.71 & 0.24 & 50 & 0.81 & 0.16 \\
 EV & - & - & 212 & 0.73 & 0.26 & 22 & 0.75 & 0.14 \\
\hline
 PS$_W$  & 8 & 200 & 1872 & 0.74 & 0.20 & 437 & 0.80 & 0.16 \\
 PS$_W$ & 5 & 453 & 1861 & 0.73 & 0.20 & 611 & 0.80 & 0.16 \\
\hline
PS$_K$ & 1 & 203 & 6807 & 0.69 & 0.22 & 566 & 0.80 & 0.16  \\
PS$_K$ & 10 & 203 & 4141 & 0.73 & 0.21 & 1054 & 0.81 & 0.17\\
\hline
PS$_K$ & 1 & 453 & 1201 & 0.70 & 0.24 & 376 & 0.81 & 0.16 \\
PS$_K$ & 2 & 453 & 1670 & 0.71 & 0.21 & 610 & 0.80 & 0.16 \\
PS$_K$ & 5 & 453 & 537 &  0.74 & 0.19 & 268 & 0.80 & 0.16 \\
\hline
\end{tabular}
\caption{The summary for anomalous exponent $\alpha$. Abbreviations: EV - biological samples (extracellular vesicles), PS$_W$ - polystyrene beads measured in Warsaw, PS$_K$ - polystyrene beads measured in Krakow, $C_0= 1\%$ - initial concentration of sample solution, $d$ - the certified diameter of a bead, $\mathcal{N}_{all}$ - number of all accepted trajectories with length $N\ge4$, $<\alpha>$ - mean anomalous exponent for all accepted trajectories, SD - standard deviation of $<\alpha>$, $\mathcal{N}_{N>50}$ - number of accepted trajectories longer than $N=50$, $<\alpha>_{N>50}$ anomalous exponent calculated only for long trajectories, SD$_{N>50}$ - standard deviation of $<\alpha>_{N>50}$.\label{tab:alpha}}
\end{table*}

\begin{figure*}
\includegraphics[width=0.9\textwidth]{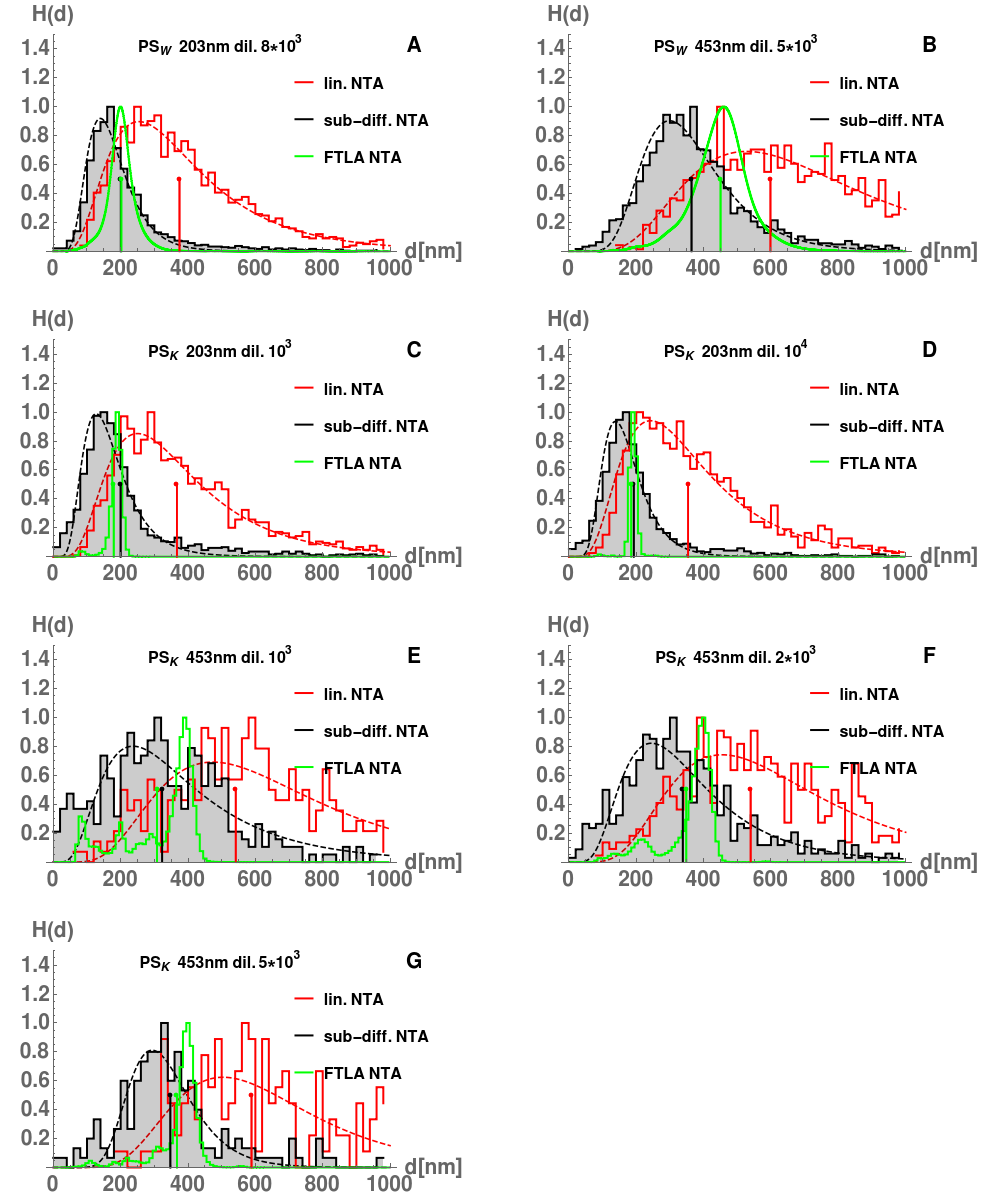}
\caption{Histograms of diameter $d$ for PS beads at different dilutions (dil.), measured with NTA. Each panel represents the size distribution obtained with the normal diffusion model (red solid line), logarithmic sub-diffusion (black solid line) and with FTLA algorithm (green solid). Dashed lines show the log-normal fits applied to the data. Distributions are normalized to have the same height at the maximum. Vertical lines indicate the position of $<d>$. PS$_W$, PS$_K$ - measurements carried out, respectively, in Warsaw and Krakow.
\label{fig:beads}}
\end{figure*}

Having proposed the criteria of data selection and truncation, we address the central issue of this letter, which is the sub-diffusive dynamics of EVs. All the selected trajectories were fitted with the three following models:
\begin{align}
&\textrm{MSD}(n)=4D_1t+C^2 \label{eq:mod1}\\
&\textrm{MSD}(n)=4D_\alpha t^{\alpha}+C^2 \label{eq:mod2}\\
&\textrm{MSD}(n)=4\frac{D_1}{v}\ln (1+vt)+C^2 \label{eq:mod3}
\end{align}
where the additional constant $C^2$ accounts for the localization noise as advocated by Berglund\cite{bib:berglund}. In Fig. \ref{fig:examples}A the exemplary fits obtained with all the three models are shown for weak and strongly sub-diffusive trajectories. While for normal diffusion they perform comparably well, it is evident that for strong sub-diffusion the linear model visibly deviates from the data. While the localization noise is able to partially remedy the quality of fits, it is still insufficient, leading to the underestimated value of $D_1$, and, in turn, to the overestimated $d$. It should be denoted that the sub-diffusive models predict extremely low level of the localization noise. 

The power-low model provides a conventional test for anomalous diffusion. Fig. \ref{fig:alphas} and Fig. \ref{fig:b_alphas} show the histograms of the anomalous exponents $\alpha$ for EVs and PS beads, respectively. The numerical results are also summarized in the Table \ref{tab:alpha}. In general, the distributions of $\alpha$ calculated for all the accepted trajectories (Fig. \ref{fig:alphas}A and Fig. \ref{fig:b_alphas}A), have a mono-peaked shape with a maximum at $<\alpha>\simeq0.69-0.74$ and the standard deviation of approximately $0.19-0.26$. This is common for all samples, i.e. both PS beads and EVs, which shows the ubiquity of the sub-diffusive behavior in NTA. However, this data include a huge number of short trajectories with low predictive power\cite{bib:saxton}, so to minimized their influence we also plot the same distributions, but only for the long ($N>50$) trajectories  (Fig. \ref{fig:alphas}B and Fig. \ref{fig:b_alphas}B). While this slightly shifts $<\alpha>$ to 0.75-0.81, the presence of sub-diffusion clearly persists. Further increase in the minimal length of admitted trajectories do not lead to a significant growth in $<\alpha>$. The experiments on the bigger beads also suggest that there is a dependence between $<\alpha>$ and concentration, i.e. as the dilution grows by 5 times, $<\alpha>$ increases from 0.7 to 0.74 (see Table \ref{tab:alpha}). This is in agreement with our conjecture that the sub-diffusion is at least partially caused by the crowding in a sample, though the effect is rather weak.

The drift of particles was also analyzed.  The mean increments $<\Delta x_n>$ and $<\Delta y_n>$ (see Fig. \ref{fig:examples}B) were fitted with the power law model $\propto t^\gamma$ to test their linearity. In Fig. \ref{fig:alphas}C and Fig. \ref{fig:b_alphas}C the histograms of $\gamma$ are shown. A distinct peak at $\gamma\simeq1$ indicate that, indeed, the drift was approximately constant during the observation. The additional peak at $\gamma=0$ was associated with the particles experiencing no flow, i.e. $<\Delta\vec{r}_n>=0$. Since the majority of particles were carried by an approximately constant drift, it is possible that they were systematically conveyed into the denser regions, in agreement with the assumptions of our logarithmic model.

\begin{figure}
\includegraphics[width=0.98\columnwidth]{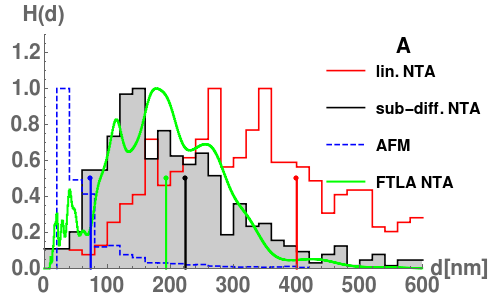}
\includegraphics[width=0.98\columnwidth]{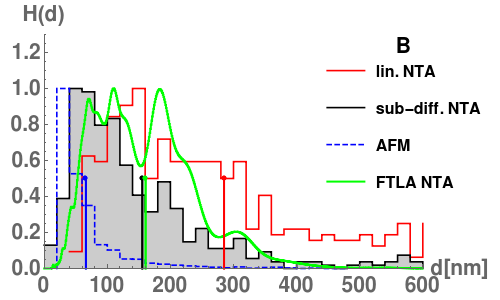}
\includegraphics[width=0.98\columnwidth]{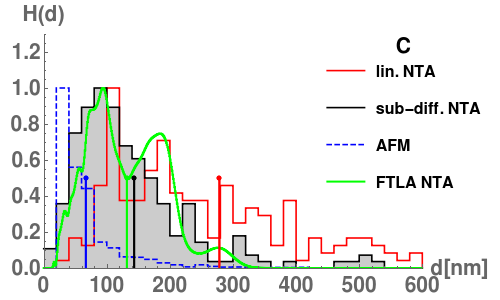}
\caption{The histograms of EVs diameter $d$ for three donors (\textbf{A} - P1,\textbf{B} - P2, \textbf{C} - P3) obtained for 3 different models. Shaded black solid line - the logarithmic sub-diffusive model, red solid line - the normal (linear) diffusion model, green line - FTLA algorithm. For comparison, the size distribution measured with AFM (blue dashed line), for the same samples, are shown. The vertical lines indicate the mean value for each distribution. Histograms are normalized to have the same height at the maximum.\label{fig:histograms}}
\end{figure}

\begin{table*}
\small
\begin{tabular}{c|c|c|ccccc|cccc}
\hline
 \multirow{2}{*}{ }&dil.&\multirow{2}{*}{} &\multicolumn{5}{c|}{$<d>$ $\pm$ SD[nm]} & \multicolumn{4}{c}{$d_{max}$[nm]}\\
 \cline{4-12}
 Id.&$\times$& d&\multicolumn{3}{c|}{NTA} & \multicolumn{2}{c|}{AFM} & \multicolumn{3}{c}{NTA}& AFM\\
 & $10^3$& [nm] & lin dif. & sub-dif. & \multicolumn{1}{c|}{FTLA }& $d\ge50$ & all & lin. diff & sub-diff. & FTLA & \\
\hline
P1& - & - & 400$\pm$361&223$\pm$293& 180$\pm$78 & 101$\pm$73 &74$\pm$65 & 253 &123 & - & 10\\
P2 & - &- &285$\pm$341 &155$\pm$263& 159$\pm$66&  91$\pm$61 &66$\pm$54 & 117 & 54 & - & 13\\
P3 & - & -& 278$\pm$263&143$\pm$189& 130$\pm$57 &  90$\pm$56 &67$\pm$51& 129 & 65 & - & 16\\
\hline
PS$_W$  & 7 & 203 & 375$\pm$188 & 200$\pm$133 & 202$\pm$47 & - & - & 254 & 140 & 199 & -\\
PS$_W$  & 5 & 453 & 598$\pm$200 & 364$\pm$162 & 450$\pm$86 & - & -& 527 & 299 & 459 & - \\
\hline
PS$_K$ & 1 & 203 & 367$\pm$272 & 200$\pm$158 & 180$\pm$25 & - & -& 250 & 127 & 185 & - \\
PS$_K$ & 10 & 203 & 354$\pm$479& 192$\pm$137 & 185$\pm$19 & - & - & 242 & 138 & 185 & -\\
\hline
PS$_K$ & 1 & 453 & 541$\pm$209 & 323$\pm$190 &307$\pm$112 & - &- &473& 237 & 385 & -\\
PS$_K$ & 2 & 453 & 539$\pm$207 & 338$\pm$191 & 348$\pm$82 &- &- & 455 & 244  & 395  &-\\
PS$_K$ & 5 & 453 & 589$\pm$200 & 348$\pm$174 & 367$\pm$72  & - & - & 504 & 293 & 395 &-\\
\hline
\end{tabular}
\caption{Size determination in the NTA and AFM measurements of EVs and PS beads of different nominal size $d$ and at different dilutions (dil.). The NTA data were analysed with the normal diffusion model, sub-diffusion model and the FTLA algorithm. For AFM the mean size $<d>$ was calculated for both: entire dataset (column \emph{all}) and for the particles with $d\ge50$nm (for direct comparison with optical methods). The table summarizes $<d>$, standard deviation SD of $d$ and the position of the distribution maximum, $d_{max}$. All parameters were determined by fitting the $H(d)$ histograms (see Fig. \ref{fig:beads} and \ref{fig:histograms}) with the log-normal function (except for FTLA, which was provided by the NanoSight software). \label{tab:d} }
\end{table*}

Another step is to compare the efficiency of size prediction between the normal diffusion model, logarithmic sub-diffusion and FTLA algorithm. As explained before, the power-law model cannot be used for this purpose. The mono-disperse solutions of PS beads with 203 nm and 453 nm diameters and at several different concentrations were used as the test samples. In this paragraph we will refer to Warsaw measurements as PS$_W$ and to experiments in Krakow as PS$_K$. The size histograms obtained for each sample are shown in the Fig. \ref{fig:beads} and the numerical results are summarized in Tab. \ref{tab:d}. In general, the distributions of diameters obtained with the normal and logarithmic diffusion models are significantly asymmetric and can be efficiently fitted with the log-normal function. Thus, one can consider the maximum and the average of this distribution as the two possible measures of the particle size. On the other hand, FTLA usually predicts the distribution dominated by a single, mostly symmetric peak. Let us consider the 203 nm beads first (Fig. \ref{fig:beads}A, C and D). FTLA recovers this size almost perfectly (202 nm) for PS$_W$ and with 91\% accuracy (180 and 185 nm, depending on the concentration) for PS$_K$. Conversely, the normal diffusion model performs very poorly in both series, predicting the mean size $<d>$ which is twice over-sized. The maximum of these distributions is also shifted by $\simeq$+25\% in comparison to the actual size. Similarly, the maximum of the logarithmic sub-diffusive model indicates the size which is underestimated by 30-40\%. However, there is a remarkable agreement between the expected size and $<d>$ in the sub-diffusive model. For PS$_W$ sub-diffusive model indicates 200 nm (98\% accuracy) and for PS$_K$ it predicts 200 and 192 nm. These last two results are even more accurate than the FTLA predictions. The situation is much different for the bigger particles, with 453 nm diameter. FTLA works almost perfectly (Fig. \ref{fig:beads}B) in PS$_W$ measurements, predicting 450 nm. However, in PS$_K$ series the artefacts appear, decreasing $<d>$ to 307-367 nm. If the position of the main peak is used, the prediction improves to 385-395 nm. This difference in FTLA performance between PS$_W$ and PS$_K$ is astonishing, as the measurement conditions and sample concentrations in both cases are comparable. However, two different versions of NanoSight software are involved in the analysis, which might be the reason. The sub-diffusive model proved less accurate than FTLA, resulting in $<d>=323-348$ nm. Surprisingly, the best estimate is provided by the normal diffusion model, when the maximum of the distribution is considered. In this case the error is lower than 11$\%$. In general, these results strongly suggest that the diffusion model applied to the NTA data should be chosen according to the expected particle diameter, as, apparently, the accuracy of methods changes with the particle size. Nevertheless, $<d>$ obtained from the logarithmic sub-diffusion model proves to be a very accurate measure for the smaller particles. 

As the final test, we applied the normal diffusion model, logarithmic sub-diffusion and FTLA to predict the $<d>$ of EVs. To corroborate this analysis, the samples were also examined with AFM, which provided the size distribution measured by a direct imaging method. The results for patients P1, P2 and P3 are plotted in the Fig. \ref{fig:histograms} and the numerical results are summarized in the Table \ref{tab:d}. The size distribution obtained with AFM is an asymmetric and mono-peaked function with $<d>$ equal 74, 66 and 67 nm, depending on the sample. Rejecting the particles with $d<50$nm shifts this averages to 101, 91 and 90 nm, respectively. Let us recall that, according to our discussion in the introductory section, the factor of discrepancy between NTA and AFM reads approximately 1.5-2.5. Using this value and the AFM data one can estimate to expected EVs size in NTA to be roughly equal 150-250nm. This is precisely the size range in which the sub-diffusive model is reliable. Indeed, the sub-diffusive model results in a $d$ histogram whose general shape is similar to that of the AFM distribution, with $<d>=$223, 155 and 143 nm. For P2 and P3 these results neatly fall into the expected range of discrepancy, while for P1 the difference is by the factor of $\simeq 3.0$, which is still close to our expectations. On the other hand, the normal diffusion model leads to a distribution with an appropriate shape, but predicts $<d>$ which is 4.0-5.5 times greater than $<d>$ from AFM. Finally, FTLA provides $<d>$ which is comparable to the sub-diffusive results, but the shape of the distribution is much different. In particular, FTLA predicts a few maxima of a similar height that weakly coincide with the structure of the normal diffusion distribution. Supposedly, this is the manifestation of the problems with FTLA in the highly poly-disperse samples, i.e. it tries to map the actual size distribution on a limited number of peaks. However, these peaks do not seem to be present in the AFM data. 

Summarizing the experimental part, we have shown that the sub-diffusive behavior occurs in both artificial and biological samples and at a range of concentrations. The diffusion model applied to the MSD data proves to have a critical impact on the particles size measured with NTA. The experiments were performed on two different instruments, thus we can suspect that the issue is not inherent to a specific experimental setup. The method of data selection and the logarithmic sub-diffusive model which we propose seem to be acceptable tools to improve the analysis of NTA data in the biological context. We have shown that for the small particles ($d\simeq200$nm) the sub-diffusive predictions are on par with the state-of-the-art FTLA methods. However, our model does not introduce the spurious manipulation of data that can cause artefacts in the poly-disperse samples. Nevertheless, the idea behind FTLA is sound and this approach might be improved in the future. Especially, combining the FTLA-type corrections with the theory of sub-diffusion might lead to a significant progress in NTA data processing for the biological applications. We have also illustrated the problem of discrepancy between AFM and NTA size prediction for EVs. Our results confirm that this difference exists and it can be minimized only by applying the analysis methods beyond the normal diffusion models.

\section{Summary}
NTA is a technique of huge potential, but its broad application to the biological research is still in development and requires the clarification of many ambiguities. In this article we analysed the problem of anomalous diffusion affecting NTA measurements in both artificial and biological poly-disperse systems. We proposed the protocol of data processing and selection that allows us to stay in agreement with the requirements of the free-diffusion model. We also introduced a simple model of sub-diffusion that can be ready-applied to retrieve the EVs size distributions. The application of the sub-diffusion models leads to the results comparable with FTLA approach, but is more reliable in the biological samples. It should be emphasized that our protocol and model can be applied in the NTA study of any poly-disperse system provided that the sub-diffusion is observed in the MSD data. The notorious discrepancy between NTA and AFM results was also attributed to the several well-recognized effects. Finally, our approach opens a way for further development of the NTA method.

\begin{acknowledgement}
Experiments have been financed by the Polish National Science Centre (NCN) grant No. 2012/07/B/NZ5/02510 (to E.S.). TEM images were performed by dr in\.{z}. Olga Wo\'znicka from the Department of Cell Biology and Imaging Institute of Zoology, Jagiellonian University. M. Majka acknowledges the National Science Center, Poland for the grant support (2014/13/B/ST2/02014).
\end{acknowledgement}

\end{document}